\documentstyle[eqsecnum,psfig,aps]{revtex}
\tightenlines
\makeatletter
\def\setcaption#1{\def\@captype{#1}}
\makeatother

\begin{document}

\title{Neutrino-induced upward stopping muons in Super-Kamiokande}

\date{\today}

\maketitle

%
%

{\center \large The Super-Kamiokande Collaboration\\}

\begin{center}
\newcounter{foots}
Y.Fukuda$^a$, K.Ishihara$^a$, Y.Itow$^a$,
T.Kajita$^a$, J.Kameda$^a$, S.Kasuga$^a$, K.Kobayashi$^a$, Y.Kobayashi$^a$, 
Y.Koshio$^a$,   
M.Miura$^a$
% ...
%
\addtocounter{foots}{1}
\addtocounter{foots}{1} 
J.S.George$^{x,\fnsymbol{foots}}$, A.L.Stachyra$^x$,
R.J.Wilkes$^x$, K.K.Young$^{x,\dagger}$

\footnotesize \it

$^a$Institute for Cosmic Ray Research, University of Tokyo, Tanashi,
Tokyo 188-8502, Japan\\
% ...
$^x$Department of Physics, University of Washington,    
Seattle, WA 98195-1560, USA    \\
\end{center}

\newpage

\begin{abstract}
  A total of 137 upward stopping muons of minimum energy 1.6~GeV are
  observed by Super-Kamiokande during 516 detector live days.  The
  measured muon flux is $0.39\pm 0.04 {\rm(stat.)}  \pm 0.02
  {\rm(syst.)}\times10^{-13} {\rm{cm^{-2}s^{-1}sr^{-1}}}$ compared to an
  expected flux of $0.73\pm 0.16{\rm(theo.)} \times
  10^{-13}{\rm{cm^{-2}s^{-1}sr^{-1}}}$.  Using our previously-published
  measurement of the upward through-going muon flux, we calculate the
  stopping/through-going flux ratio $\mathcal{R}$, which has less
  theoretical uncertainty. The measured value of $\mathcal{R}$ =0.22$\pm
  0.02{\rm(stat.)} \pm 0.01 {\rm(syst.)}$ is significantly smaller than
  the value $0.37^{+0.05}_{-0.04}{\rm(theo.)}$ expected using the best
  theoretical information (the probability that the measured
  $\mathcal{R}$ is a statistical fluctuation below the expected value is
  0.39\%).  A simultaneous fitting to zenith angle distributions of
  upward stopping and through-going muons gives a result which is
  consistent with the hypothesis of neutrino oscillations with the
  parameters $\sin^{2} 2\theta >0.7$ and
  1.5$\times$10$^{-3}$\(<\)$\Delta
  m^2$\(<\)1.5$\times$10$^{-2}$~eV$^{2}$ at 90\% confidence level,
  providing a confirmation of the observation of neutrino oscillations
  by Super-Kamiokande using the contained atmospheric neutrino events.
\end{abstract}





\vspace{5mm}

\noindent
{\bf keywords:} {\it neutrino~oscillations, Super-Kamiokande,
  upward~stopping~muons, upward~throughgoing~muons,
  atmospheric~muon~neutrinos, cosmic~rays}

\vspace{5mm}

\newpage

\begin{figure}[thbp]
\psfig{file=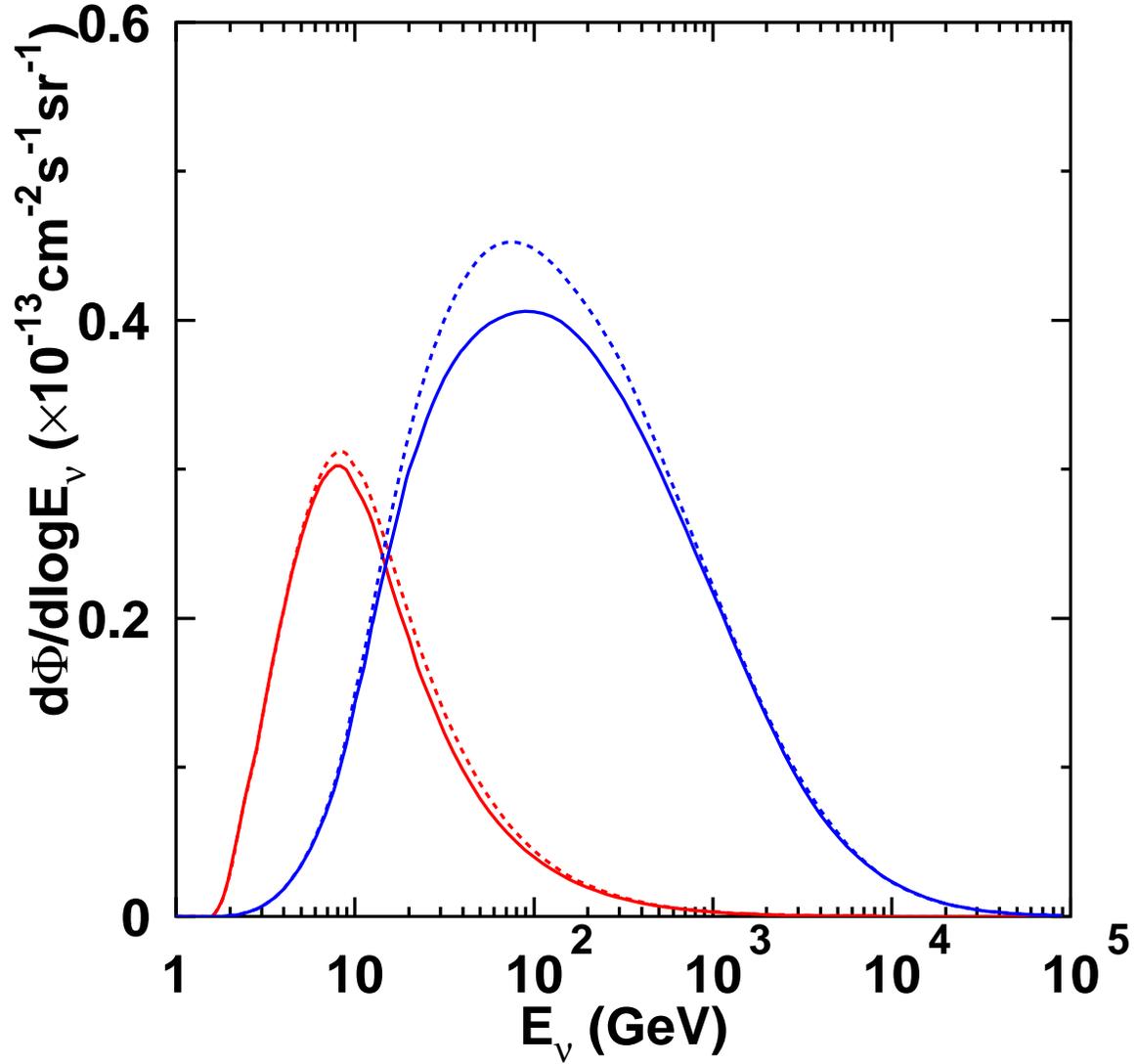,width=6.5in}
  \caption{The energy spectra of the parent neutrinos of upward stopping
    muons (left) and upward through-going muons (right).  The dashed
    lines are the result of using the Bartol input fluxes, and the solid
    lines are from Honda's calculations.}
  \label{fig:spectra}
\end{figure}

Energetic atmospheric $\nu_{\mu}$ or $\bar{\nu}_{\mu}$ interact with the
rock surrounding the Super-Kamiokande (``Super-K'') detector and produce
muons via weak interactions.  For downward-going particles, the rock
overburden is insufficient to prevent cosmic-ray muons from overwhelming
any neutrino induced muons, but upward-going muons are $\nu_\mu$ or
$\bar{\nu}_\mu$ induced because the entire thickness of the earth
shields the detector.  Muons energetic enough to cross the entire
detector are defined as ``upward through-going muons'', and have been
discussed in a previous article~\cite{skupthrumu}.  The typical energy
of the parent neutrinos is approximately 100~GeV.  Upward-going muons
that stop in the detector are defined as ``upward stopping muons'', and
come from parent atmospheric neutrinos with a typical energy of about
10~GeV.  These energy spectra are shown in Fig.~\ref{fig:spectra}.
Neutrinos arriving vertically from beneath the detector travel roughly
13,000~km from their point of production, while those coming from near
the horizon originate only $\sim$500~km away. Thus, observation of
upward-going muons provides a relatively pure sample of muon neutrinos,
with a wide range of path lengths, allowing tests of possible $\nu_\mu$
disappearance due to flavor neutrino oscillations~\cite{maki}.

For atmospheric neutrinos with interaction vertices inside the
detector's fiducial volume (referred to as ``contained'' events),
Super-K measures a low $\nu_\mu/\nu_e$ ratio~\cite{sksubgev} and clearly
observes a strong zenith angle dependence of the $\nu_\mu$
events~\cite{skmultigev}.  This muon neutrino disappearance~\cite{skosc}
is consistent with $\nu_\mu \leftrightarrow \nu_\tau$ oscillations,
while the up-down symmetry seen in the $\nu_e$ events rules out
significant $\nu_e$ appearance, in agreement with recent results from
the CHOOZ experiment~\cite{chooz}.  The $\nu_\mu \leftrightarrow
\nu_\tau$ oscillation hypothesis suggested to explain the Super-K
contained event results is also consistent with anomalous upward
through-going muon zenith angle distributions observed by
Kamiokande~\cite{kamupthrumu}, MACRO~\cite{macro} and
Super-Kamiokande~\cite{skupthrumu}.  The upward stopping muons are the
remaining class of atmospheric neutrino events with which Super-K can
test the oscillation hypothesis.  It should be noted here that the
parent neutrino energies of upward stopping muons are similar to those
of multi-GeV fully-contained (FC) and partially-contained (PC) events
discussed in~\cite{skmultigev}.  Thus, a similar deficit of upward
stopping muons to that of upward-going multi-GeV FC and PC events is
expected if the oscillation hypothesis is the cause of the multi-GeV FC
and PC event deficit.  The main difference between the multi-GeV FC and
PC event sample and the upward stopping muon sample is that roughly 80\%
of the upward stopping muons are generated in the surrounding rock,
which has different neutrino interaction cross-sections and systematic
uncertainties than does water.

The Super-K detector is a 50~kton cylindrical water Cherenkov detector.
To reduce the cosmic-ray muon background, the detector was constructed
$\sim$1000~m (2700~m.w.e.) underground at the Kamioka Observatory,
Institute for Cosmic Ray Research, the University of Tokyo, in the
Kamioka-Mozumi mine, Japan.  The detector is divided by an optical
barrier instrumented with photomultiplier tubes (``PMT''s) into a
cylindrical primary detector region (the Inner Detector, or ``ID'') and
a surrounding shell of water (the Outer Detector, or ``OD'') serving as
a cosmic-ray veto counter.  Details of the detector and general data
reduction procedures can be found in reference~\cite{sksubgev}. The data
used in this analysis were taken from Apr.~1996 to Jan.~1998,
corresponding to 516 days of detector livetime.

An upward-going muon is defined as a track that appears to enter the
detector from the rock, and reconstructs as traveling in the upward
direction.  Thus, PMT activity in the OD at the muon's entrance point is
required.  The total cosmic-ray muon rate at Super-K is 2.2~Hz, of which
a few percent are stopping muons, and the great majority are
downward-going.  The trigger efficiency for a muon entering the ID with
momentum more than 200~MeV/c is $\sim$100\% for all zenith angles.

Muons which leave entrance signal clusters in the OD with no
corresponding exit cluster are regarded as stopping.  A neutrino
interaction inside the water of the OD also produces such a signal, so
some fraction ($\sim 20\%$) of the upward-going stopping muon sample is
estimated to originate in the OD rather than the rock according to
simulations.  This effect has been accounted for in the expected flux
calculations.

Stopping muons with track length \(>\) 7~m ($\sim$1.6~GeV) in the ID are
selected for further analysis.  The stopping muon track length is
determined by calculating the muon momentum from a photoelectron count
in the same way as in the contained event analysis~\cite{sksubgev}.
This cut eliminates short tracks that are very close to the PMTs and
thus are hard to reconstruct, provides an energy threshold, and also
eliminates nearly all contamination of the upward stopping muon signal
by pions.  These pions are photoproduced by cosmic-ray muons outside the
detector and are observed to have a soft spectrum~\cite{up-pions}, such
that residual contamination upward-going pions meeting the 7~m track
length requirement is estimated to be \(<\) 1\%.  The nominal detector
effective area for upward-going muons with a track length \(>\) 7~m in
the ID is $\sim$1200~m$^2$.

137 upward stopping muons which satisfy a $\cos\Theta<0$ cut are found,
where $\Theta$ is the zenith angle of the muon track, with
$\cos\Theta=-1$ corresponding to vertically upward-going events.
  
Details of the muon track reconstruction method and data reduction
algorithm are similar to those of through-going muons described
elsewhere~\cite{skupthrumu}.  The total detection efficiency for the
complete data reduction process for upward stopping muons is estimated
by a Monte Carlo simulation to be \(>\)98\% over $-1<\cos\Theta<0$. The
validity of this Monte Carlo estimate is in turn checked with real
cosmic-ray downward stopping muons, taking advantage of the up/down
symmetry of the detector geometry.

\begin{figure}[thbp]
\psfig{file=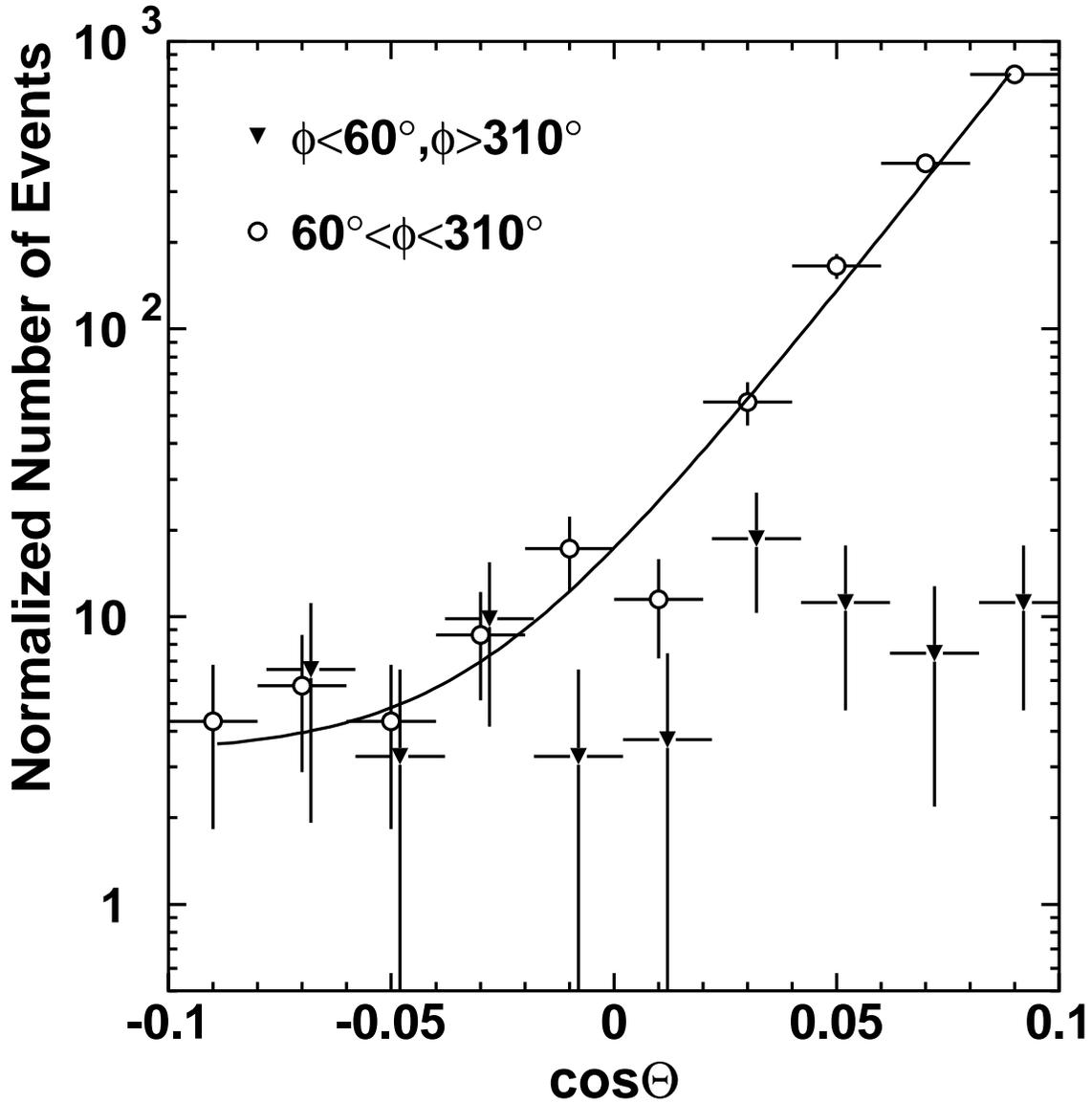,width=6.5in}
  \caption{Zenith angle distribution 
    of stopping muons near the horizon observed by Super-K.  Filled
    triangles (open circles) indicate events coming from direction where
    the rock overburden is thick $>15000$ m.w.e. (shallow $>5000$ m.w.e.).
    The two distributions are normalized to a common azimuth angle range
    (0 - 360 degrees).  The solid line is the zenith fit to the shallow
    rock events used to estimate the cosmic-ray muon background
    contamination in the $-0.2<\cos\Theta<0$ bin.}
  \label{fig:backg_fit}
\end{figure}

\begin{figure}[thbp]
\psfig{file=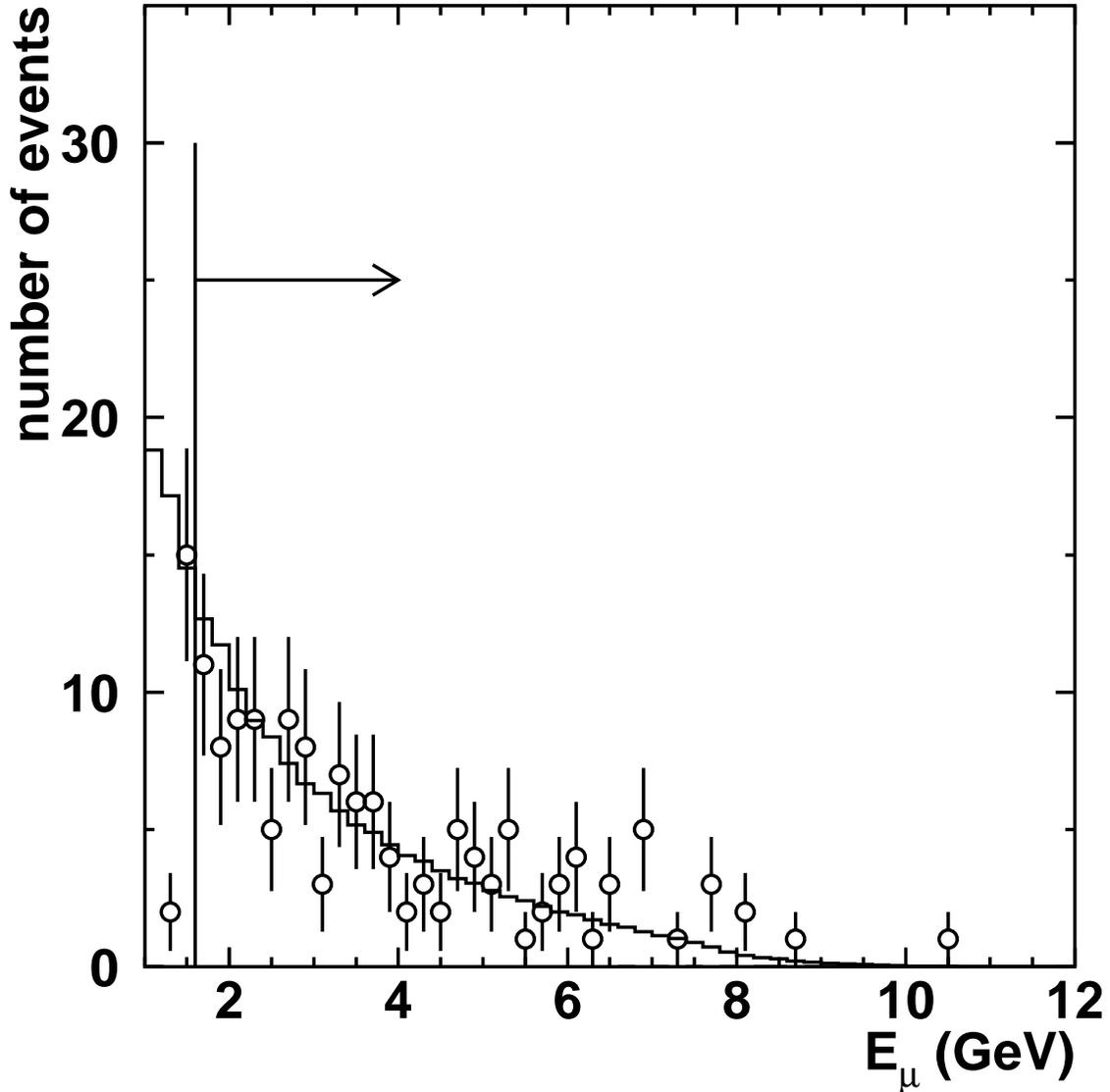,width=6.5in}
  \caption{The energy distribution of upward stopping
    muons, illustrating the freedom of the data both from threshold
    effects due to the 7~m (1.6~GeV) muon pathlength cut and from
    upward-scattered low energy stopping cosmic-ray muons.  The solid
    histogram is the normalized distribution expected from the
    no-oscillations hypothesis. The normalization is made so that the
    integrated number of expected events with muon energy $>$ 1.6~GeV
    (the vertical line) corresponds to the observed number of events
    (=137).  Preselection is applied at muon energy $\sim$1.4~GeV.}
  \label{fig:Eupstop}
\end{figure}

Because of multiple Coulomb scattering and finite muon fitter angular
resolution ($\sim1^\circ$~\cite{skupthrumu}), some of the abundant
downward-going cosmic-ray stopping muon background may be reconstructed
with $\cos\Theta<0$ and contaminate the neutrino induced upward stopping
muon sample.  Figure~\ref{fig:backg_fit} illustrates the estimation of
this contamination by extrapolation of the downward-going zenith angle
distribution to the flat neutrino-induced signal near the horizon.  This
background falls exponentially with decreasing $\cos\Theta$, the
contribution to apparent upward stopping muons is estimated to be
10.8$\pm4.2$ events in the zenith angle bin with $\cos\Theta$ between
-0.2 and 0.  This is a larger contamination than was present in the
upward through-going muon sample due to the lower muon energies (and
correspondingly larger multiple scattering) of the stopping muon sample,
and is more uncertain due to lower statistics.  As an independent
crosscheck of the background contamination, the energy spectrum of the
observed stopping muons is shown in Fig.~\ref{fig:Eupstop}. That shape is
consistent with the expected and no significant contamination is
observed at the 1.6 GeV muon energy analysis threshold.

To analytically calculate the expected upward stopping (and
through-going as well) muon flux, employed is the combination of the
Bartol atmospheric neutrino flux model~\cite{bartol}, a neutrino
interaction model composed of quasi-elastic (QE) scattering~\cite{qe} +
single-pion (1$\pi$) production~\cite{sp} + deep inelastic (DIS)
scattering multi-pion production based on the parton distribution
functions (PDF) of GRV94~\cite{grv94} with the additional kinematic
constraint of $W>1.4$~GeV/c$^2$ (where $W$ is the invariant mass of the
hadronic recoil system), and Lohmann's muon energy loss formula
in standard rock~\cite{lohmann}.

This expected flux is compared to three other analytic calculations to
estimate the model-dependent uncertainties of the expected muon flux.
The other flux calculations use various pairwise permutations of the
Honda flux~\cite{honda} or the atmospheric neutrino flux model
calculated by the Bartol group~\cite{bartol}, the GRV94DIS PDF or the
CTEQ3M~\cite{cteq}PDF.  This comparison yields $\pm$10\% of difference
in the overall flux and -2\% to +1\% for the bin-by-bin shape difference
in the zenith-angle distribution.  The shape difference is due mostly to
differences in the input flux models.

The expected muon flux $\Phi^{stop}_{theo}$ resulting from the above
calculation is $0.73\pm 0.16 \times10^{-13}{\rm{cm^{-2}s^{-1}sr^{-1}}}$
($\cos\Theta < 0$), where the estimated theoretical uncertainties are
described in Table~\ref{systematictable_norm_th}.  The dominant error
comes from the overall normalization uncertainty in the neutrino flux,
which is estimated to be approximately
$\pm20$\%~\cite{bartol,honda,frati} above several GeV.

Given the detector live time $T$, the effective area for upward stopping
muons $S(\Theta)$, and the detection efficiency $\varepsilon(\Theta)$,
the upward stopping muon flux is calculated using:
\begin{displaymath}
  \Phi^{stop}=\sum^{N}_{j=1} \frac{1}{\varepsilon(\Theta_j)}
  \cdot\frac{1}{S(\Theta_j)\,2\pi}\cdot\frac{1}{T}
\end{displaymath}
\noindent
where the index $j$ runs over observed events, $2\pi$ is the total solid
angle covered by the detector for upward stopping muons, and $N$ is the
total number of observed muon events (137).  Subsequently, we subtract
the cosmic-ray muon contamination (10.8 events) from the most horizontal
bin (-0.2\(<\)cos$\Theta$\(<\)0).  The resulting observed upward
stopping muon flux is: $\Phi^{stop}=0.39\pm0.04 {\rm(stat.)} \pm 0.02
{\rm (sys.)}\times10^{-13} {\rm{cm^{-2}s^{-1}sr^{-1}}}$.  If instead of
subtracting this background, we simply omit upward stopping muons coming
from the thin-rock region~\cite{skupthrumu} in the most horizontal
cos$\Theta$ bin (-0.1\(<\)cos$\Theta$\(<\)0 and 60$^{\circ}$ \(<\)
$\phi$ \(<\)310$^{\circ}$), the calculated flux differs from the
background-subtracted flux by -1.1\%.  Systematic errors in the
experimental measurement are summarized in
Table~\ref{systematictable_norm_ex}.

\begin{figure}[thbp]
\psfig{file=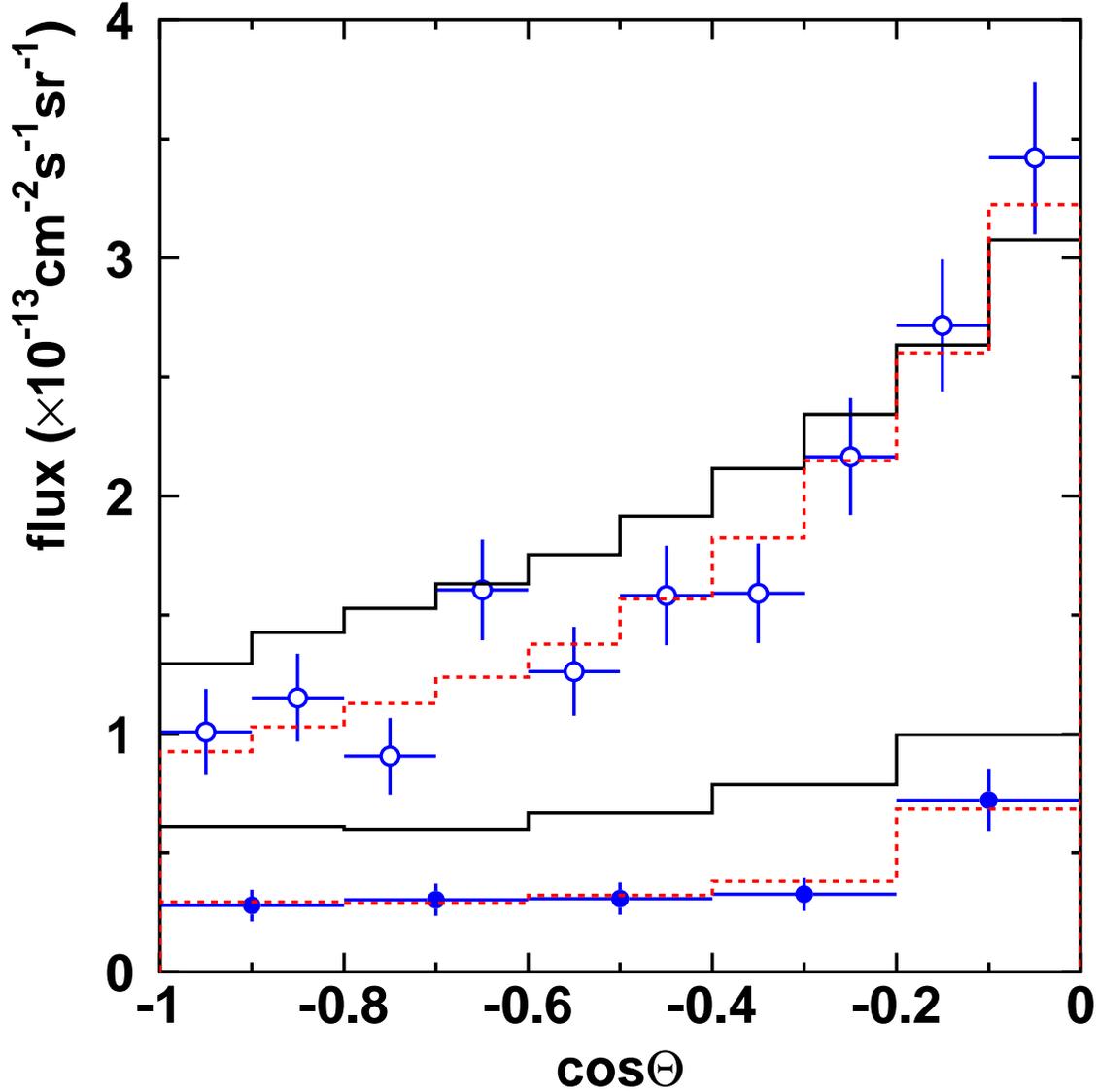,width=6.5in}
  \caption{Upward stopping (filled circles) and 
    through-going (open circles) muon fluxes observed in Super-K as a
    function of zenith angle.  The error bars indicate uncorrelated
    experimental systematic plus statistical errors added in quadrature.
    The lower (upper) solid histograms show the expected upward stopping
    (through-going) muon flux based on the Bartol neutrino flux without
    oscillation.  Also shown as lower (upper) dashed histograms are the
    expected stopping (through-going) muon flux assuming the best fit
    parameters of the combined analysis in the physical region at
    $(\sin^2 2\theta, \Delta m^2)=(1.0, 3.9\times10^{-3}{\rm{eV}}^{2})$,
    $\alpha=+0.061$ and $\beta$=-0.083 for the
    $\nu_{\mu}\leftrightarrow\nu_{\tau}$ oscillation case.}
  \label{fig:flux_zen}
\end{figure}

The flux as a function of zenith angle, $(d\Phi^{stop}/d\Omega)$, is
shown in Fig.~\ref{fig:flux_zen}.  Due to limited statistics for
stopping muons, 5 angular bins are used instead of the 10 bins used for
Super-K through-going muons.  With the present statistics, the shape of
the normalized distribution is consistent with the no-oscillations
hypothesis ($\chi^{2}$/d.o.f. = 4.1/4 corresponding to 39\%
probability).  However, the overall flux of upward stopping muons
observed is substantially depressed from that expected, as this shape
comparison is made after multiplying the expected flux by a free-running
overall normalization factor (1+$\alpha$) whose best fit value is
$\alpha=-51\%$ calculated by a $\chi^{2}$ shape fit .  This can be
compared with an observed no-oscillations $\alpha = -14\%$ in the
through-going muon sample~\cite{skupthrumu}.

\begin{figure}[thbp]
\psfig{file=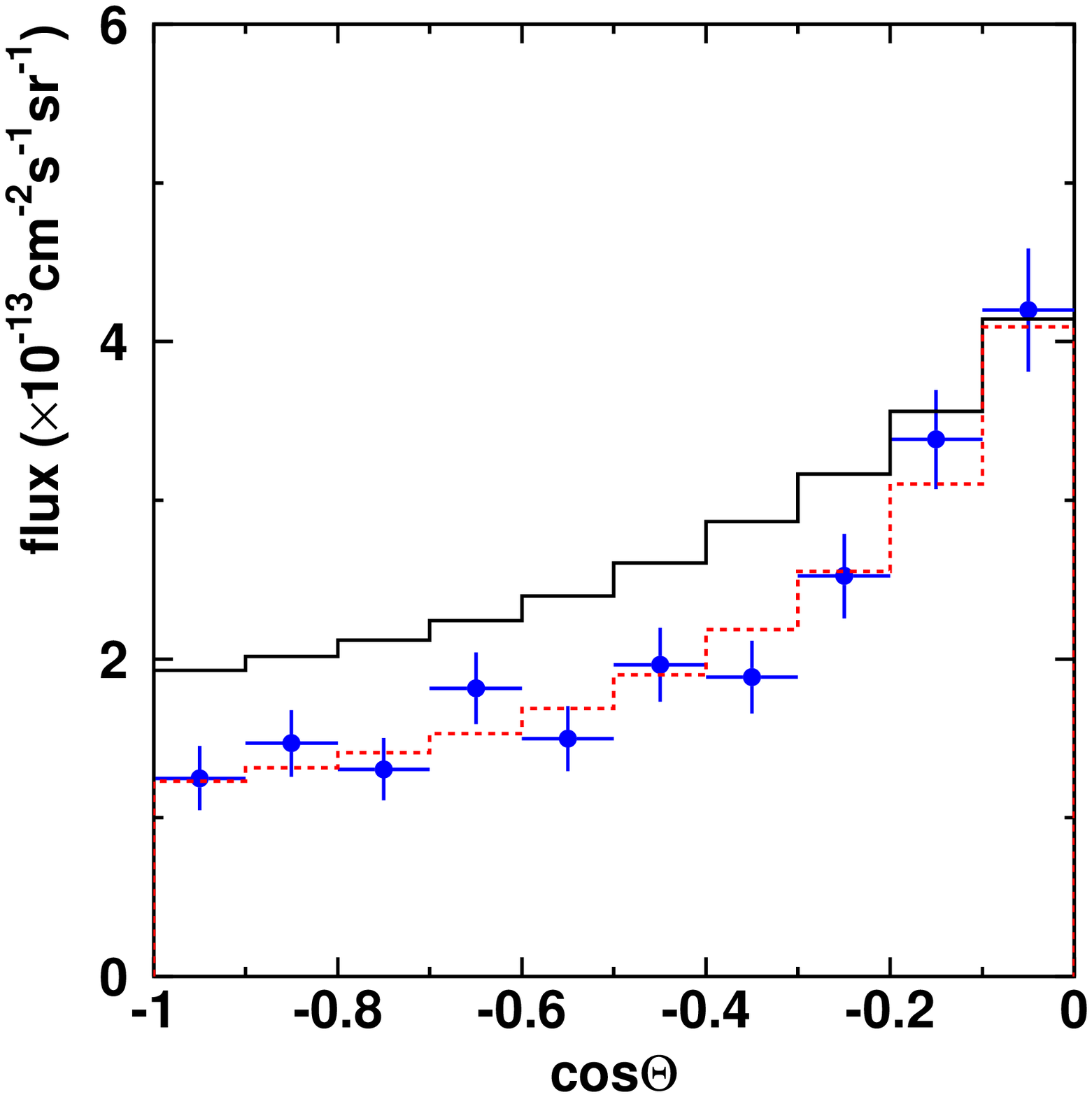,width=6.5in}
  \caption{Upward stopping+through-going muon flux (filled circles)
    observed in Super-K as a function of zenith angle.  The error bars
    indicate uncorrelated experimental systematic plus statistical
    errors added in quadrature.  The solid histograms show the expected
    flux based on the Bartol neutrino flux without oscillation.  Also
    shown as dashed histograms are the expected flux assuming the best
    fit parameters of the combined analysis in the physical region at
    $(\sin^2 2\theta, \Delta m^2)=(1.0, 3.9\times10^{-3}{\rm{eV}}^{2})$,
    $\alpha=+0.061$ and $\beta$=-0.083 for the
    $\nu_{\mu}\leftrightarrow\nu_{\tau}$ oscillation case.}
  \label{fig:upmu_zen}
\end{figure}

Adding the upward through-going flux~\cite{skupthrumu} $\Phi^{thru}$, we
obtain a detector-independent total upward-going muon flux
$\Phi^{stop+thru}$ of $2.13\pm0.08 {\rm(stat.)}\pm0.03
{\rm(sys.)}\times10^{-13} {\rm{cm^{-2}s^{-1}sr^{-1}}}$ with muon energy
\(>\) 1.6~GeV.  The zenith angle distribution of the summed flux,
$(d\Phi^{stop+thru}/d\Omega)$, is shown in Fig.~\ref{fig:upmu_zen}.

\begin{figure}[thbp]
\psfig{file=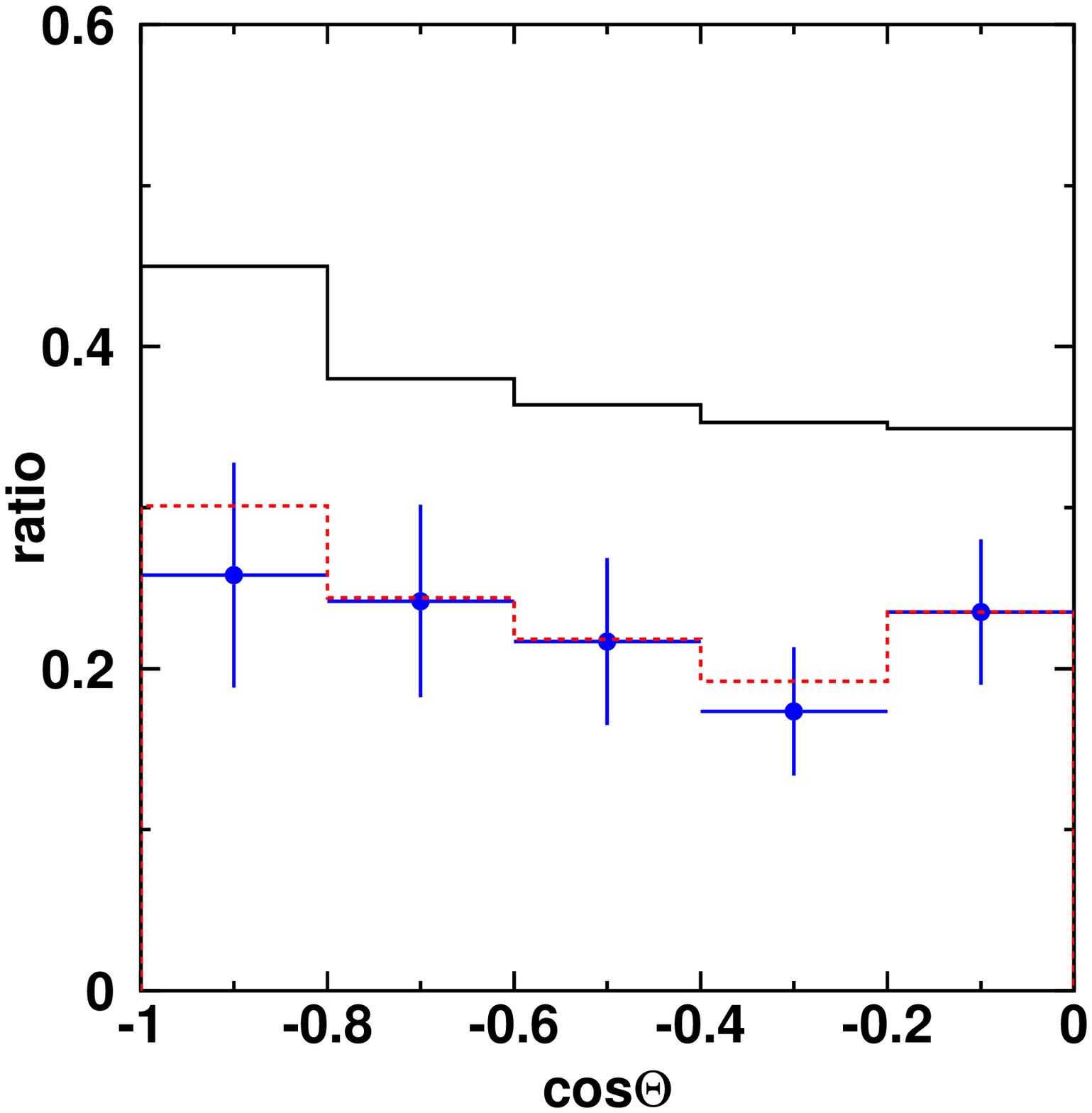,width=6.5in}
  \caption{Upward stopping/through-going muon flux ratio 
    (filled circles) observed in Super-K as a function of zenith angle.
    The error bars indicate uncorrelated experimental systematic plus
    statistical errors added in quadrature.  The solid histograms show
    the expected the ratio based on the Bartol neutrino flux without
    oscillation.  Also shown as dashed histograms are the expected ratio
    assuming the best fit parameters of the combined analysis in the
    physical region at $(\sin^2 2\theta, \Delta m^2)=(1.0,
    3.9\times10^{-3}{\rm{eV}}^{2})$, $\alpha=+0.061$ and $\beta$=-0.083
    for the $\nu_{\mu}\leftrightarrow\nu_{\tau}$ oscillation case.}
  \label{fig:rat_zen}
\end{figure}

A more useful physical quantity than the absolute flux for probing
neutrino oscillations is the stopping/through-going flux ratio
${\mathcal{R}}=\Phi^{stop}/\Phi^{thru}$, which cancels much of the large
($\sim20\%$) uncertainty in the neutrino flux normalization and the
neutrino interaction cross sections\cite{imbstop}.  The systematic
theoretical and experimental uncertainties in $\mathcal{R}$ are
summarized in Table~\ref{systematictable_th} and
Table~\ref{systematictable_ex}, respectively.  The measured
${\mathcal{R}}$ is $0.22\pm0.02 {\rm(stat.)}\pm0.01 {\rm(sys.)}$, while
the expected ${\mathcal{R}}_{theo}$ is
$0.37^{+0.05}_{-0.04}{\rm(theo.)}$.  The probability that the observed
ratio could fluctuate this far below the expectation is 0.39\%.
The zenith angle distribution of the ratio $\mathcal{R}$ is shown in
Fig.~\ref{fig:rat_zen}.

\begin{figure}[thbp]
\psfig{file=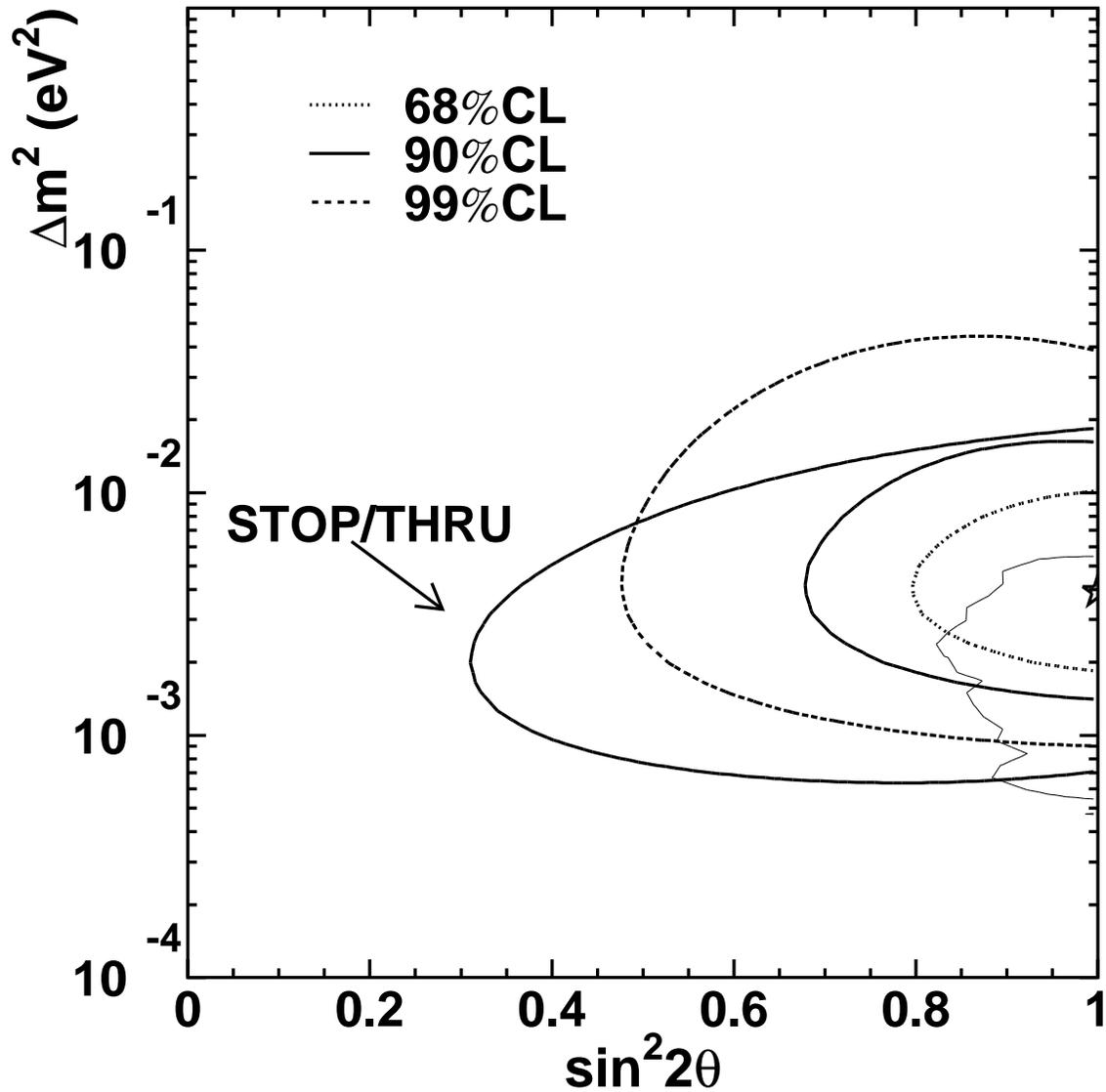,width=6.5in}
  \caption{The allowed region contours at 68\% (dotted contour), 
    90\% (thick solid), and 99\% (dashed) C.L. obtained by the combined
    analysis of Super-K upward stopping and through-going muons drawn on
    the ($\sin^{2}2\theta$,$\Delta{m}^{2}$) plane for
    $\nu_{\mu}\leftrightarrow\nu_{\tau}$ oscillations.  The star
    indicates the best fit point at $(\sin^2 2\theta, \Delta m^2)=(1.0,
    3.9\times 10^{-3}{\rm{eV}^2})$ in the physical region.  The allowed
    region contour indicated by solid thick labelled line with
    "STOP/THRU" is made based on the Super-K stopping/through-going muon
    ratio alone at 90\% C.L.  Also shown is the allowed region contour
    (the remaining solid thin line) at 90\% C.L. by the Super-K
    contained event analysis.  The allowed regions are to the right of
    the contours.}
  \label{fig:allowed}
\end{figure}

Using these data, we derive probability contours on the neutrino
oscillation parameter $(sin^{2} 2 \theta,\Delta m^{2})$ plane for the
$\nu_{\mu}\leftrightarrow\nu_{\tau}$ oscillation hypothesis, as shown in
Fig.~\ref{fig:allowed}, based on a $\chi^2$ defined by:
\begin{eqnarray*}
  \chi^{2}&=&
        {\min_{\beta}}\left[
        \sum_{i=1}^{5(stop)}\left(
        \frac{{\mathcal{R}}^i-
         {\mathcal{R}}^i_{theo}(sin^{2} 2 \theta,\Delta m^{2}) \times(1+\beta)}
        {\sqrt{\sigma_{{\mathcal{R}}^{i}_{stat}}^{2}+
        \sigma_{{\mathcal{R}}^{i}_{sys}}^{2}}}
        \right)^{2}\right.\\
        & &\left.\hspace{10mm}
        +\left(\frac{\beta}{\sigma_{\beta}}\right)^{2}\right]
\end{eqnarray*}

\noindent 
where ${\mathcal{R}}^i$ is the observed ratio in the $i$-th cos$\Theta$
bin, $\sigma_{{\mathcal{R}}^{i}_{stat}}$ the experimental statistical
error, $\sigma_{{\mathcal{R}}^{i}_{sys}}$ the bin-by-bin uncorrelated
systematic error ($\simeq$2\%) estimated by adding uncorrelated
theoretical and experimental systematic errors in
Table~\ref{systematictable_th} and Table~\ref{systematictable_ex} in
quadrature, and ${\mathcal{R}}^i_{theo}$ the expected ratio.  The
$\beta$ for ${\mathcal{R}}^i_{theo}$ represents a
stopping-to-throughgoing relative muon flux normalization factor with
error $\sigma_{\beta}$=14\% estimated by the correlated theoretical and
experimental systematic errors in Table~\ref{systematictable_th} and
Table~\ref{systematictable_ex} added in quadrature.

The allowed region thus obtained in Fig.~\ref{fig:allowed} is in good
agreement with that found in the Super-K contained event
analysis~\cite{skosc}.  The minimum $\chi^{2}$ locations on the $\Delta
m^{2}-\sin^{2}2\theta$ plane and corresponding parameter values for
various conditions are listed in Table~\ref{tab:chisqminima}.  As the
minimum $\chi^{2}$ lies in the unphysical region, the contour is drawn
according to the prescription for bounded physical regions given in
Ref.~\cite{pdg}.  Because the $\chi^{2}$ surface has a rather broad
minimum, the specific best-fit oscillation parameter values cited are of
less importance than the contours shown in Fig.~\ref{fig:allowed}.  If
we replace the Bartol neutrino flux~\cite{bartol} by the Honda's
flux~\cite{honda} and/or the GRV94DIS parton distribution
functions~\cite{grv94} by CTEQ3M~\cite{cteq}, the allowed region
contours do not change significantly.

Finally, we performed an oscillation analysis which use all the
upward-going muon information available by simultaneously fitting to the
upward through-going and stopping muon zenith angle distributions.  In
this analysis, $\chi^{2}$ is defined by the sum over 10
upward-throughgoing and 5 upward-stopping muon zenith angle bins:

\begin{eqnarray*}
  \chi^{2}={\min_{\alpha, \beta}}
\left[ 
\sum_{i=1}^{10(thru)} \left(
  \frac{\left(\frac{d\Phi}{d\Omega}\right)^{i}-
    (1+\alpha)\left(\frac{d\Phi}{d\Omega}\right)_{theo}^{i}(sin^{2}
2\theta,\Delta m^{2})}
      {\sqrt{(\sigma^{i}_{stat})^{2}
        +(\sigma^{i}_{sys})^{2}}}\right)^{2}
\right.\\+
\sum_{i=1}^{5(stop)} \left(
  \frac{\left(\frac{d\Phi}{d\Omega}\right)^{i}-
    (1+\alpha)(1+\beta)
        \left(\frac{d\Phi}{d\Omega}\right)_{theo}^{i}
            (sin^{2} 2\theta,\Delta m^{2})}
  {\sqrt{(\sigma^{i}_{stat})^{2}
        +(\sigma^{i}_{sys})^{2}}}\right)^{2}\\
\left.
+ \left(\frac{\alpha}{\sigma_{\alpha}}\right)^{2}
+ \left(\frac{\beta}{\sigma_{\beta}}\right)^{2}
\right]
\end{eqnarray*}
\noindent
where $\left(\frac{d\Phi}{d\Omega}\right)^{i}$ is the observed muon flux
in the $i$-th cos$\Theta$ bin, $\sigma_{stat}^{i}$ the experimental
statistical error, $\sigma_{sys}^{i}$ ($\sim$2 to $\sim$4\%) the
bin-by-bin uncorrelated theoretical and experimental systematic errors
in Table~\ref{systematictable_norm_th} and
Table~\ref{systematictable_norm_ex} added in quadrature for $stop$ or
from Ref.~\cite{skupthrumu} for $thru$ ,
$\left(\frac{d\Phi}{d\Omega}\right)_{theo}^{i}$ the expected muon flux,
$\alpha$ the running overall flux normalization factor with error
$\sigma_{\alpha}$ of 22\% estimated by adding in quadrature the
correlated theoretical systematic errors in
Table~\ref{systematictable_norm_th}, $\beta$ the running
stopping-to-throughgoing relative muon flux normalization factor with
error $\sigma_{\beta}$=14\% estimated by the correlated theoretical and
experimental systematic errors in Table~\ref{systematictable_th} and
Table~\ref{systematictable_ex} added in quadrature.  The results are
shown in Fig.~\ref{fig:allowed} and are consistent with those from the
Super-K contained event analysis.  As is shown in
Table~\ref{tab:chisqminima}, the minimum $\chi^{2}$ falls in the
unphysical region so the contours are drawn according to the
prescription for bounded physical regions given in Ref.~\cite{pdg}.

The contamination due to $\nu_{e}$ charged current interactions and
neutral current interactions in the upward-going muons is estimated to
be $<$~1\% by a Monte Carlo simulation and is neglected in these
analyses.  The contribution of possible $\nu_\tau$ interactions in the
rock below to the upward stopping muon flux is suppressed by branching
ratios and kinematics to less than a few percent, and is also neglected.

The observed overall upward stopping muon flux alone and its zenith
angle distribution do not conflict with the expected no-oscillations values
significantly within the present statistical and systematic errors.
However, in order to simultaneously explain both the stopping and
through-going upward muon fluxes using both an analysis based on the
stop/through ratio $\mathcal{R}$ and a combined fit of upward stopping
and through-going muon zenith angle data, the no-oscillations
theoretical expectations reproduce the observed fluxes poorly at the
0.39\% C.L. (stop/thru ratio averaged over zenith angle) and 0.87\%
C.L, respectively, while the $\nu_{\mu}\leftrightarrow\nu_{\tau}$
oscillation assumption is consistent with observations.  This result
supports the evidence for neutrino oscillations given by the analysis of
the contained and through-going muon atmospheric neutrino events by
Super-K.

We gratefully acknowledge the cooperation of the Kamioka Mining and
Smelting Company.  The Super-Kamiokande experiment has been built and
operated from funding by the Japanese Ministry of Education, Science,
Sports and Culture, and the United States Department of Energy.

\begin{table}[tbhp]
  \caption{List of theoretical uncertainties in upward stopping muon flux calculation.}
  \begin{tabular}{cl|c}
    \multicolumn{2}{c|}{Error source} & Error  \\
    \hline
    \multicolumn{2}{l|}{typical $\nu$ flux normalization}
                       & $\pm$20\%\tablenotemark[1] \\ \hline
    \multicolumn{2}{l|}{Choice of $\nu$ flux/PDF} & \\
                       & overall flux & $\pm$10\%\tablenotemark[1] \\
                       & bin by bin & -2\% to +1\% \tablenotemark[2]\\
  \end{tabular}
  \tablenotetext[1]{Theoretical cos$\Theta$ bin-by-bin correlated uncertainty}
  \tablenotetext[2]{Theoretical cos$\Theta$ bin-by-bin 
                    uncorrelated uncertainty}
  \label{systematictable_norm_th}
\end{table}

\begin{table}[thbp]
  \caption{List of experimental systematic errors in upward stopping muon flux
    measurement.}
  \begin{tabular}{l|l}
    Error source & Error  \\
    \hline
    7~m track length cut                 & $\pm$5\%\tablenotemark[1]\\
    Live time                           & $\pm$1\%\tablenotemark[1]\\
    Reconstruction efficiency             & $\sim \pm1$\%\tablenotemark[2]\\

  \end{tabular}
  \tablenotetext[1]{Experimental cos$\Theta$ bin-by-bin 
                    correlated systematic error}
  \tablenotetext[2]{Experimental cos$\Theta$ bin-by-bin 
                    uncorrelated systematic error}
  \tablenotetext{Note that the uncertainty of the background subtraction
                 is included in the statistical error.}
  \label{systematictable_norm_ex}
\end{table}

\begin{table}[tbhp]
  \caption{List of theoretical uncertainties in stopping/through-going
    muon flux ratio.}
  \begin{tabular}{cl|c}
    \multicolumn{2}{c|}{Error source} & Error  \\
    \hline
    \multicolumn{2}{l|}{Primary spectral index $\pm$0.05}
                       & $\pm$13\%\tablenotemark[1] \\
    \multicolumn{2}{l|}{Choice of $\nu$ flux} & \\
                        & overall ratio & $\pm$1\%\tablenotemark[1] \\
                        & bin by bin & -1\% to +2\%\tablenotemark[2] \\
    \multicolumn{2}{l|}{$\nu$ cross section \tablenotemark[3]}
                       & $\pm$ 4\%\tablenotemark[1] \\

  \end{tabular}
  \tablenotetext[1]{Theoretical cos$\Theta$ bin-by-bin correlated uncertainty}
  \tablenotetext[2]{Theoretical cos$\Theta$ bin-by-bin
                    uncorrelated uncertainty}
\tablenotetext[3]{QE, 1$\pi$ and DIS cross sections are changed
  independently by $\pm$15\%, and the effect of this on $\mathcal{R}$
  is noted here.}
  \label{systematictable_th}
\end{table}

\begin{table}[thbp]
  \caption{List of experimental systematic errors in 
    stopping/through-going muon flux ratio.}
\begin{tabular}{c|c}
    Error source & Error  \\
    \hline
    7~m track length cut                 & $\pm$5\%
\tablenotemark[1]\\
    Live time                           & $\pm$1\%
\tablenotemark[1]\\
    stop/thru misidentification         & $\sim \pm$1\%
\tablenotemark[1]\\
    Reconstruction efficiency           & $\sim \pm$1\%
\tablenotemark[2]\\
\end{tabular}
\tablenotetext[1]{Experimental cos$\Theta$-bin-by-bin correlated
  error}
\tablenotetext[2]{Experimental cos$\Theta$-bin-by-bin uncorrelated
  error}

  \label{systematictable_ex}
\end{table}

\begin{table}[tp]
  \caption{Summary of fit results. See text for definitions of $\alpha$
    and $\beta$.}
  \begin{tabular}{l|c|c|c|c|c|r}
    Case & $\Delta m^{2}$ &  $sin^{2}2\theta$  & $\alpha$ & $\beta$& $\chi^{2}_{MIN}$/DOF & Probability\\
    \hline
Flux vs. zenith angle (Fig.~\ref{fig:flux_zen}): & - & - & -0.51 & - & 4.1/4 &
39\% \\
\hline
$\mathcal{R}$ analysis (Fig.~\ref{fig:rat_zen}):&&&&&&\\
Fit in the  physical region & $3.1\times 10^{-3}{\rm{eV}^2}$ & 1.00 & 
- & -0.084 & 0.76/3 & 86\% \\
Fit over all space & $3.9\times 10^{-3}{\rm{eV}^2}$ & 1.19 & - & -0.003 & 0.30/3 &
96\% \\
No oscillations & - & - & - & -0.35 & 8.4/5 & 13\% \\
\hline
(stop+through) cos($\Theta$) analysis (Fig.~\ref{fig:allowed}):&&&&&&\\
Fit in the physical region & $3.9\times 10^{-3}{\rm{eV}^2}$ & 1.00 &
0.061 & -0.083 & 8.8/13 & 79\% \\
Fit over all space & $3.9\times 10^{-3}{\rm{eV}^2}$ & 1.13 & 0.10 & -0.012 & 8.2/13 & 83\% \\
No oscillations & - & - & -0.16 & -0.36 & 31.0/15 & 0.87\% \\
  \end{tabular}
  \label{tab:chisqminima}
\end{table}

\end{document}